\documentclass[aps,prc,twocolumn,nofootinbib,superscriptaddress]{revtex4-2}

\usepackage[utf8]{inputenc}
\usepackage[pdfencoding=auto,psdextra]{hyperref}
\usepackage{amsmath}
\usepackage{amssymb}
\usepackage{mathtools}
\usepackage{physics}
\usepackage{graphicx}
\usepackage{url}
\usepackage{blindtext}
\usepackage{units}
\usepackage[dvipsnames]{xcolor}
\usepackage{pifont}
\usepackage{simplewick}
\usepackage{wasysym}
\usepackage{placeins}
\usepackage{comment}
\usepackage{xifthen}
\usepackage{anyfontsize}
\usepackage{xcolor}
\usepackage{bigstrut}

\usepackage[labelfont={small},font=small,subrefformat=parens,caption=false]{subfig}
\usepackage{physics}
\usepackage{dcolumn}

\graphicspath{{./figures/}}

\usepackage{environ}
\NewEnviron{topic}{}{}

\usepackage[disable]{todonotes}
\newcommand{\newabbreviation}[3]{\newcounter{#1}\expandafter\newcommand\csname#1\endcsname[1][]{\ifthenelse{\equal{##1}{abreviate}}{#2}{\ifthenelse{\equal{##1}{fullname}}{#3}{\ifthenelse{\equal{##1}{explain}}{#3 (#2)\stepcounter{#1}}{\ifthenelse{\value{#1}=0}{#3##1 (#2##1)\stepcounter{#1}}{#2##1}}}}}}

\newabbreviation{HO}{\text{HO}}{harmonic oscillator}
\newabbreviation{NCSM}{\text{NCSM}}{no-core shell model}
\newabbreviation{MBSE}{\text{MBSE}}{many-body Schrödinger equation}
\newabbreviation{LEC}{\text{LEC}}{low-energy constant}
\newabbreviation{LO}{\text{LO}}{leading order}
\newabbreviation{NLO}{\text{NLO}}{next-to-leading order}
\newabbreviation{NNLO}{\text{N2LO}}{next-to-next-to-leading order}
\newabbreviation{CoM}{\text{CM}}{center-of-mass}
\newabbreviation{threeNF}{\text{3NF}}{three-nucleon force}
\newabbreviation{twoNF}{\text{2NF}}{two-nucleon force}
\newabbreviation{EFT}{EFT}{effective field theory}
\newabbreviation{chiEFT}{$\chi$EFT}{chiral effective field theory}
\newabbreviation{NZME}{NZME}{non-zero matrix elements}
\newabbreviation{SD}{\text{SD}}{Slater determinant}
\newabbreviation{EC}{\text{EC}}{eigenvector continuation}
\newabbreviation{PPD}{\text{PPD}}{posterior predictive distribution}
\newabbreviation{PDF}{\text{PDF}}{probability density function}
\newabbreviation{MAP}{\text{MAP}}{maximum a posteriori}
\newabbreviation{MLE}{\text{MLE}}{maximum likelihood estimation}
\newabbreviation{MCMC}{\text{MCMC}}{Markov Chain Monte Carlo}
\newabbreviation{iid}{\text{i.i.d.{}}}{independent and identically distributed} 

{}  

\DeclareMathOperator{\pr}{pr} 
\newcommand{\given}{\,|\,}  

\newcommand{\cbar}{\ensuremath{\bar{c}}}

\newcommand{\inputvec}{\boldsymbol}
\newcommand{\obs}[1][]{\ensuremath{y_\textup{#1}}}
\newcommand{\obsset}[1][]{\ensuremath{\inputvec{\obs}_\textup{#1}}}
\newcommand{\err}[1][]{\ensuremath{\delta y_\textup{#1}}}
\newcommand{\errset}[1][]{\ensuremath{\inputvec{\err}_\textup{#1}}}
\newcommand{\data}[1][]{\ensuremath{\mathcal{D}_\textup{#1}}}

\newcommand{\obsc}[1][]{\ensuremath{c_\textup{#1}}}
\newcommand{\obscset}[1][]{\ensuremath{\inputvec{\obsc}_\textup{#1}}}
\newcommand{\cov}[1][]{\ensuremath{\Sigma_\textup{#1}}}
\newcommand{\stddev}[1][]{\ensuremath{\sigma_\textup{#1}}}
\newcommand{\corr}[1][]{\ensuremath{R_\textup{#1}}}

\newcommand{\obsth}{\ensuremath{\obs[th]}}       

\newcommand{\obsMB}{\ensuremath{\obs[NCSM]}}       
\newcommand{\obsMBset}{\ensuremath{\obsset[NCSM]}}       
\newcommand{\obsMBinf}{\ensuremath{\obs[NCSM,$\infty$]}}       
\newcommand{\obsref}{\ensuremath{\obs[ref]}}

\newcommand{\errEFTset}{\ensuremath{\errset[EFT]}}
\newcommand{\errMB}{\ensuremath{\err[NCSM]}}       
\newcommand{\errMBset}{\ensuremath{\errset[NCSM]}}
\newcommand{\errem}{\ensuremath{\err[em]}}       
\newcommand{\erremset}{\ensuremath{\errset[em]}}

\newcommand{\JNCSM}{\textsc{JupiterNCSM}}
\newcommand{\nsopt}{\textsc{nsopt}}
\newcommand{\pantoine}{\textsc{pAntoine}}
\newcommand{\NCSD}{\textsc{NCSD}}

\newcommand{\hw}{\ensuremath{\hbar\Omega}}
\newcommand{\Nmax}{\ensuremath{N_\text{max}}}

\newcommand{\NN}{\ensuremath{\text{NN}}}
\newcommand{\piN}{\ensuremath{\pi\text{N}}}
\newcommand{\NNN}{\ensuremath{\text{3N}}}
\newcommand{\onePECT}{\ensuremath{\text{3NF},1\pi\operatorname{-}\text{ct}}} 
\newcommand{\twoPE}{\ensuremath{\text{3NF},2\pi}}
\newcommand{\threeCT}{\ensuremath{\text{3NF},\text{ct}}}

\newcommand{\lecs}{\ensuremath{\vec{a}}}
\newcommand{\lecsNN}{\ensuremath{\lecs_{\NN}}}
\newcommand{\lecspiN}{\ensuremath{\lecs_{\piN}}}
\newcommand{\lecsNNN}{\ensuremath{\lecs_{\NNN}}}
\newcommand{\cD}{\ensuremath{c_D}}
\newcommand{\cE}{\ensuremath{c_E}}
\newcommand{\cDk}{\ensuremath{c_{D,k}}}
\newcommand{\cEk}{\ensuremath{c_{E,k}}}

  \def\nuc#1#2{\relax\ifmmode{}^{#1}{\protect\text{#2}}\else${}^{#1}$#2\fi}
  \def\itnuc#1#2{\setbox\@tempboxa=\hbox{\scriptsize\it #1}
    \def\@tempa{{}^{\box\@tempboxa}\!\protect\text{\it #2}}\relax
    \ifmmode \@tempa \else $\@tempa$\fi}

\newcommand{\prA}{\ensuremath{\pr_\mathrm{full}(\cD,\cE)}}
\newcommand{\prB}{\ensuremath{\pr_\mathrm{fix}^{E_{3,4}}(\cD,\cE)}}
\newcommand{\prC}{\ensuremath{\pr_\mathrm{fix}^\mathrm{all}(\cD,\cE)}}

\begin{document}

\title{Fast \& rigorous predictions for $A=6$ nuclei with Bayesian
  posterior sampling}

\author{T.\ Djärv}
\email{djarv@chalmers.se}
\affiliation{Department of Physics, Chalmers University of Technology, SE-412 96 G\"oteborg, Sweden}
\author{A.\ Ekstr\"om}
\affiliation{Department of Physics, Chalmers University of Technology, SE-412 96 G\"oteborg, Sweden}
\author{C.\ Forss\'en}
\email{christian.forssen@chalmers.se}
\affiliation{Department of Physics, Chalmers University of Technology, SE-412 96 G\"oteborg, Sweden}
\author{H.\ T.\ Johansson}
\affiliation{Department of Physics, Chalmers University of Technology, SE-412 96 G\"oteborg, Sweden}

\date{\today}

\begin{abstract}
   We make \emph{ab initio} predictions for the $A=6$ nuclear level
   scheme based on two- and three-nucleon interactions up to
   next-to-next-to-leading order in chiral effective field theory
   ($\chi$EFT). We utilize eigenvector continuation and Bayesian
   methods to quantify uncertainties stemming from the many-body
   method, the $\chi$EFT truncation, and the low-energy constants of
   the nuclear interaction.
   The construction and validation of emulators is made possible via
   the development of \JNCSM{}---a new $M$-scheme no-core shell model
   code that uses on-the-fly Hamiltonian matrix construction
   for efficient, single-node computations up to $\Nmax=10$ for
   \nuc{6}{Li}.
   We find a slight underbinding of \nuc{6}{He} and \nuc{6}{Li},
   although consistent with experimental data given our theoretical
   error bars. As a result of incorporating a correlated
   $\chi$EFT-truncation errors we find more precise predictions
   (smaller error bars) for separation energies: $S_d\left(
   \nuc{6}{Li}\right) = 0.89 \pm 0.44$~MeV, $S_{2n}\left(
   \nuc{6}{He}\right) = 0.20 \pm 0.60$~MeV, and for the beta decay
   $Q$-value: $Q_{\beta^-}\left( \nuc{6}{He}\right) = 3.71 \pm
   0.65$~MeV. We conclude that our error bars can potentially be
   reduced further by extending the model space used by \JNCSM{}.

\end{abstract}

\maketitle

\section{Introduction} \label{sec:intro}
%
The use of \emph{ab initio} many-body solvers and \EFT{} descriptions
of nuclear interactions promises to deliver rigorous uncertainty
quantification in the theoretical modeling of low-energy nuclear
observables. The key advantage 
is that this combination makes it possible to systematically
quantify the magnitude of errors made when approximating the solution
of the many-body problem and when modeling the nuclear interaction to
a finite EFT order.

A Bayesian framework is propitious for carrying out such an
uncertainty quantification program.  
Best practices for using Bayesian methods in the context of \EFT{}
descriptions have been presented in, e.g.,
Refs.~\cite{Schindler:2008fh,Furnstahl:2015rha,Wesolowski:2015fqa,Wesolowski:2018lzj,
  Melendez:2019izc,Wesolowski:2021cni}. Recently, such methods have
also been applied in the context of many-nucleon systems from
posterior sampling in the few-body sector~\cite{Kravvaris:2020lhp,
  Wesolowski:2021cni}, to estimates of \EFT{} truncation errors in
light nuclei~\cite{Maris:2020qne} and infinite nuclear
matter~\cite{Drischler:2020hwi,Drischler:2020yad}. However, the full
incorporation of pertinent sources of uncertainties in computationally
expensive many-body predictions is challenging. In this paper we
present both new computational technology and well-motivated
statistical models for relevant error terms to take first steps
towards a comprehensive analysis of uncertainties in \emph{ab initio}
nuclear structure theory.

The goals of this paper can be summarized as follows:
\begin{enumerate}
\item{Use Bayesian methods to demonstrate the propagation of
  parametric (statistical) uncertainties of the \threeNF{} \LEC[s]{}
  \cD{} and \cE{} to $A=6$-body systems using a newly developed
  \NCSM{} many-body solver.}
\item{Construct and sample statistical models for both (many-body)
    method and (\EFT{}) model uncertainties to obtain the \PPD{} for a
    set of many-body observables.} 
\item{Introduce a new open-source, \NCSM{} computer
    code---\JNCSM{}~\cite{djarv:jupiterNCSM}---that is is designed to
    solve relatively large many-body problems on single compute
nodes such that it becomes possible to efficiently explore a large set of
Hamiltonian parametrizations.} 
\item{Construct and demonstrate the accuracy of \EC{}
    emulators~\cite{Frame:2017fah,Konig:2019adq} for 
    $A=6$ \NCSM{} observables.}
\end{enumerate}

We start in Sec.~\ref{sec:interaction} with a description of a recent
Bayesian inference analysis~\cite{Wesolowski:2021cni} of nuclear
interaction parameters conditioned on calibration data in the two- and
few-nucleon sector. The many-body calculations are presented in
Sec.~\ref{sec:mb} where we also introduce the \JNCSM{} code and
construct \EC{} emulators. Our uncertainty quantification for nuclear
observables is summarized in \PPD[s]{} as discussed in
Sec.~\ref{sec:ppd}. Here we also develop statistical models to link
our theoretical models with reality in order to make final
predictions. We conclude in Sec.~\ref{sec:summary} with a summary and
outlook.

\section{Interaction model with statistical constraints} \label{sec:interaction}
%
\textcite{Wesolowski:2021cni} recently inferred posterior \PDF[s]{} for
the interaction \LEC[s]{} in \chiEFT{} at \LO{}, \NLO{}, and \NNLO{}
of the chiral expansion. In particular, they analyzed the constraints
on the leading \threeNF{} that appears at \NNLO{} from several
few-nucleon observables: the \nuc{3}{H} and \nuc{4}{He} binding
energies, the \nuc{4}{He} charge radius and the Gamow-Teller matrix
element extracted from tritium $\beta$-decay. Henceforth we label this
set of model calibration observables as \data{}. The parameter
estimation considered both experimental and computational
uncertainties as well as the model discrepancy that originates in the
truncation of the \chiEFT{} Hamiltonian. The latter was included using
a statistical error model~\cite{Wesolowski:2018lzj,Melendez:2017phj}
that relies on two parameters: the EFT expansion parameter $Q$ and the
scale $\bar{c}$ of observable coefficients. The latter governs the magnitude
of relative corrections at each \chiEFT{} order to a single observable
according to
\begin{equation}
  \obsth = \obsref \sum_n c_n Q^n,    
  \label{eq:obsSum}
\end{equation}
where it is also assumed that all coefficients $c_n$ are \iid{} random
variables following a normal distribution with zero mean and variance
$\cbar^2$: $c_n \sim \mathcal{N}(0,\cbar^2)$. We will revisit this
error model in Sec.~\ref{sec:errors} where we incorporate correlated
truncation errors for a vector of observable predictions.

The vector of \LEC{} parameters is collectively denoted \lecs{}. It
includes a subset of elements \lecsNN{} that governs the strengths of
the \twoNF{} contact interactions and a subset \lecsNNN{} that
parametrize the strengths of the shorter-range diagrams of the
leading \threeNF{} appearing at \NNLO{} in Weinberg power counting. At
this order \lecsNNN{} consists of \cD{} and \cE{}, the uncertainties
of which are a main focus of this paper. Starting at \NNLO{}, the
\chiEFT{} interaction also includes longer-range pion-nucleon (\piN{})
interactions parametrized by \lecspiN{}. In this work, we fix the
corresponding \LEC[s]{} at mean values determined in a Roy-Steiner
analysis of \piN{} scattering data~\cite{siemens2017}.

The analysis by \textcite{Wesolowski:2021cni} was performed with a
non-local momentum-space regulator function as in Eqs.~(5) and (6) of
Ref.~\cite{carlsson2015} with a single cutoff $\Lambda = 450$~MeV and
$n = 3$.
The \lecsNN{} \LEC[s]{} were optimized to reproduce \NN{}
scattering data, while the \piN{} \LEC[s]{} were fixed.
The mean values of the \NN{} and \piN{} \LEC[s]{}
are listed in Appendix B of Ref.~\cite{Wesolowski:2021cni}, while the
narrow Gaussian distribution of the \NN{} \LEC[s]{} (\lecsNN{}) is shown in Fig.~2
of the same paper. In the present analysis we will assume that both
\lecsNN{} and \lecspiN{} are fixed at their respective mean values
from the analysis in Ref.~\cite{Wesolowski:2021cni}.

The output from the parameter estimation performed in
Ref.~\cite{Wesolowski:2021cni} was a multi-dimensional posterior
\PDF{} that was sampled using \MCMC{} methods. Such parameter
posteriors can be used to identify correlations between \LEC[s]{} and
to propagate uncertainties to nuclear observables.

\begin{figure}
  \includegraphics[width=\columnwidth]{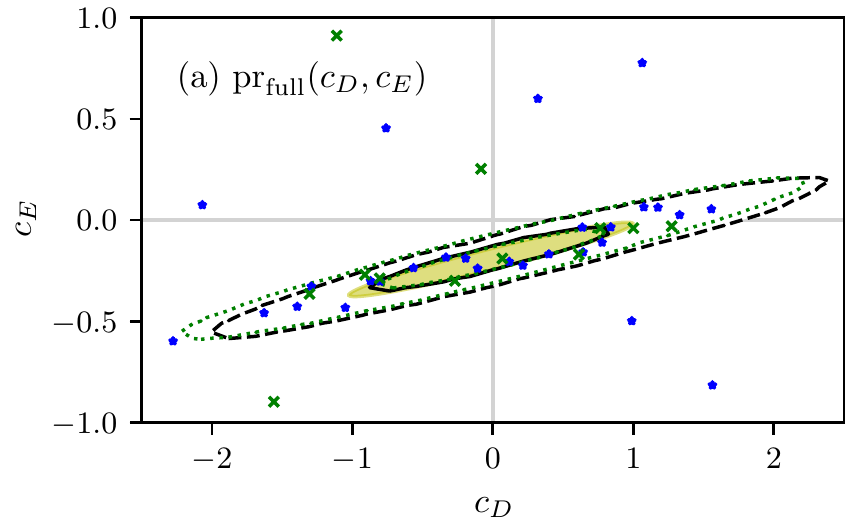}\\ 
  \includegraphics[width=\columnwidth]{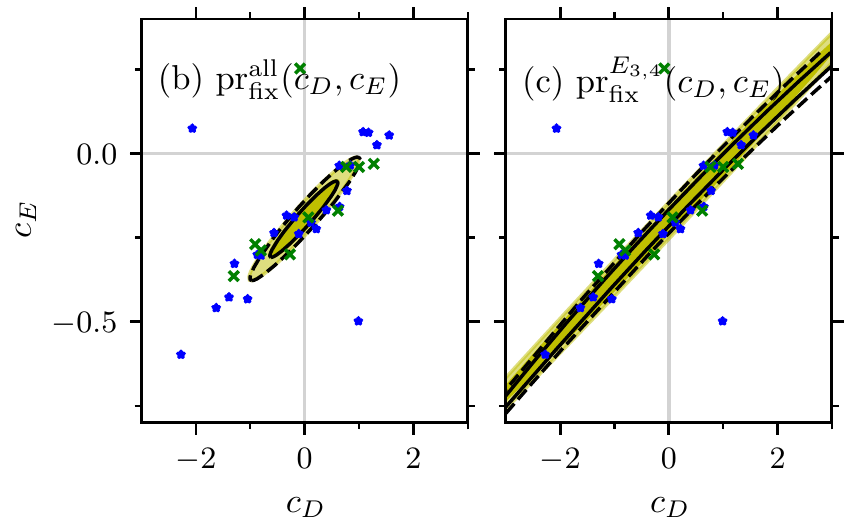} 
  \caption{Posterior pdf for \cD{} and \cE{} from
    \textcite{Wesolowski:2021cni}. The main results in this work are
    obtained with the samples from the full posterior shown in panel
    (a). For comparison we also made predictions with a posterior
    where \NN{} LECs and truncation error parameters were kept fixed,
    panels (b) and (c). The posterior shown in panel (c) was obtained
    with \nuc{3}{H} and \nuc{4}{He} binding energies as the only
    calibration data, and reveals an effective constraint in a single
    direction. The solid (dashed) iso-probability level encloses 68
    (95)\% probability mass. The yellow dark (light) ellipses correspond to
    Gaussian approximations of the sampled \PDF[s]{} and show the 68
    (95)\% probability mass region. Note that the Gaussian
    approximation is too narrow to describe the full posterior in
    panel (a). Instead, a student t distribution provides a much
    better fit~\cite{Wesolowski:2021cni} as shown by the dotted level
    curves.
    Green crosses (blue stars)
    correspond to training (validation) points for the emulators
    described in Sec.~\ref{sec:emulator}.
    \label{fig:cDcE_pdf}%
  }
\end{figure}

Marginalizing the the posterior $\pr(\lecs,\cbar^2,Q \given \data, I)$
of Ref.~\cite{Wesolowski:2021cni} over all parameters except \cD{} and
\cE{} we obtain the \PDF{}
\begin{equation}
  \prA = \int d\lecsNN d\cbar^2 dQ \pr(\lecs,\cbar^2,Q \given \data,
  I),
  \label{eq:prA}
\end{equation}
shown in Fig.~\ref{fig:cDcE_pdf}(a).

The proposition $I$ is used to implicitly subsume other known
information, such as the convergence pattern~\eqref{eq:obsSum} and the
natural scale of the \LEC[s]. Although Bayesian probability theory
only deals with conditional probabilities we sometimes suppress this
notation in favor of notational convenience.

We will also consider two alternative parameter estimation analyses
that were performed in Ref.~\cite{Wesolowski:2021cni}: The first one,
denoted \prC{}, is shown in Fig.~\ref{fig:cDcE_pdf}(b) and involved
fixing \cbar{}, $Q$ and \lecsNN{} during sampling of \cD{} and
\cE{}. It results in a more narrow distribution for \cD{} and
\cE{}. The second one, denoted \prB{}, is shown in
Fig.~\ref{fig:cDcE_pdf}(c) and was obtained with a much reduced set of
calibration data, using only the binding energies of \nuc{3}{H}
and \nuc{4}{He}. It is obvious, from visual inspection of the \PDF{}
in Fig.~\ref{fig:cDcE_pdf}(c), that these two observables are strongly
correlated such that they do not provide two independent constraints
on \cD{} and \cE{}. Also here, \cbar{}, $Q$ and \lecsNN{} were fixed
during sampling.
In this work we will use $2.5 \cdot 10^6$ (\cD{}, \cE{}) samples from \prA{} to make
predictions for the $A=6$ level scheme. For comparison we will also
make predictions with the other
two parameter \PDF[s]{} for which we use $1.0 \cdot 10^6$ samples from the
Gaussian approximations shown in Fig.~\ref{fig:cDcE_pdf}, panels (b)
and (c), respectively.
%
\section{Many-body calculations} \label{sec:mb}
%
It is a specific goal of this work to solve the \MBSE{}
\begin{equation} \label{eq:mbse}
  H(\lecs)\ket{\Psi} = E(\lecs)\ket{\Psi},
\end{equation}
for different target nuclei and with several realistic interaction
parametrizations constructed for various truncations (up to \NNLO{})
of the \chiEFT{} expansion. As indicated, each specific many-body
Hamiltonian depends on a vector of \LEC[s]{} \lecs{}.

The problem of solving Eq.~\eqref{eq:mbse} becomes significantly more
challenging at \NNLO{} with the inclusion of \threeNF{s}. The
\threeNF{} part of our Hamiltonian is parametrized by \cD{} and \cE{},
whose values are given by inferred probability distributions
$\pr(\cD,\cE)$ as described in Sec.~\ref{sec:interaction}.

Specifically, the \NNLO{} Hamiltonian consists of the intrinsic
kinetic energy plus \twoNF{} and \threeNF{} terms,
\begin{equation}
  H(\cD,\cE) = T_{\rm{int}}+V_{\twoNF}+V_{\threeNF}(\cD,\cE),
  \label{eq:hamiltonian}
\end{equation}
with
\begin{equation}
  V_{\threeNF}(\cD,\cE) = V_{\twoPE} + \cD V_{\onePECT} +\cE V_{\threeCT},
\end{equation}
where the three terms in the \threeNF{} correspond to two-pion
exchange ($\twoPE$), one-pion exchange plus contact ($\onePECT$), and
three-nucleon contact ($\threeCT$) diagrams. Note that the
$V_{\twoPE}$-term is completely determined by the fixed \LEC{} values
in \lecspiN.
For future reference we group all terms that will remain fixed in the
Hamiltonian~\eqref{eq:hamiltonian} into a constant operator \(H_0\), while the
\cD{} and \cE{} dependence enter linearly with operators \(H_1\) and
\(H_2\). That is, the full Hamiltonian is written
\begin{equation}
  H(\cD,\cE) = H_0 + \cD H_1 +\cE H_2,
\end{equation}
with
\begin{equation}
  \begin{aligned}
    H_0 =& T_{\rm{int}} + V_{\twoNF} + V_{\twoPE}\\
    H_1 =& V_{\onePECT}\\
    H_2 =& V_{\threeCT}.
  \end{aligned}
\end{equation}
%
\subsection{The \JNCSM{} code} \label{sec:jupiter}
%
To solve the \MBSE{}~\eqref{eq:mbse} we use the $M$-scheme \NCSM{} method in
which the eigenstates are expanded in a many-body \HO{} basis truncated on
the total \HO{} excitation number. Introducing the truncation
parameter \Nmax{} we have the constraint
\begin{equation}
  \sum_{i=1}^A (2n_i+l_i) - N_{\min} \leq \Nmax,
\end{equation}
where \(n_i (l_i) \) is the principal \HO{} (orbital angular momentum)
quantum number of nucleon \(i\) and $N_{\min}$ is the minimum total
excitation number ($N_{\min}=2$ for \nuc{6}{He} and \nuc{6}{Li}).
In this basis, the \MBSE{} becomes a finite matrix eigenvalue problem,
which is then solved iteratively using the Lanczos algorithm.

The \NCSM{} calculations in this work were performed with the in-house
developed \JNCSM{} code \cite{djarv:jupiterNCSM} (unless otherwise
stated). \JNCSM{} is designed to avoid storing the full Hamiltonian
matrix---a feature which makes it possible to run the code on a single compute
node. Instead, the non-zero elements are generated on the fly as
needed in the matrix-vector multiplication that lies at the heart of
the Lanczos algorithm. This multiplication is done efficiently by
precomputing the matrix representations of the \twoNF{} and \threeNF{}
interaction operators, as well as (interaction-independent)
index-lists that represent non-zero transition density elements
between the many-body basis states.

\JNCSM{} employs the proton-neutron formalism---being inspired by the code
\textsc{Antoine} by E. Caurier et al.~\cite{Caurier:1999tq,Navratil:2003ib}---with basis states being
products of \SD[s]{}, one for each nucleon species. The basis dimensions of
the $Z$-proton and $N$-neutron subspaces are much smaller than the
total dimension of the combined $A$-body basis.
This property allows to create compact index lists with one-, two-,
and three-nucleon transitions within the respective subspaces such that
allowed transitions in the many-body basis can be reconstructed.
The subspace bases are furthermore organized in blocks of fixed
energy, parity and total spin projection.

In this work we solve the \MBSE{}~\eqref{eq:mbse} with full
\threeNF{}~\eqref{eq:hamiltonian} for many different values of \cD{}
and \cE{}. However, much of the precomputed data is interaction
independent and can be generated once for each nucleus and model space
and then reused. In particular, this applies to the
generation of transition index lists and to the transformation of the
Hamiltonian terms  \(H_0\), \(H_1\) and
\(H_2\)  from $J$- to $M$-scheme. 
We restrict the calculations to model spaces with
$\Nmax \le 10$ for \nuc{6}{Li}. In this case the generation of index
lists takes 48~h, the $J$- to $M$-scheme transformation takes 16~h and
each converged Lanczos diagonalization (100 iterations) takes $\lesssim 25$~h
using a single compute node with two Intel Xeon Gold 6130 having 16 CPU cores
each and 384~GiB RAM memory.
The \nuc{6}{He} calculations could
only be performed up to \(\Nmax = 8\) due to large memory consumption
when computing the three-neutron transitions.

When running \JNCSM{} we have the choice between two different
stopping criteria for the Lanczos algorithm. Option (1) terminates the
algorithm when the difference between two subsequent iterations of the
desired eigenvalue is less than a specified tolerance $\epsilon_1$. Option (2)
terminates when the difference between two eigenvectors from subsequent Lanczos
iterations (measured as the 2-norm) reaches below a specific tolerance $\epsilon_2$. This second
option was used in this study when seeking
high-precision eigenvectors for the construction of \EC{} emulators
(see Sec.~\ref{sec:emulator}).
%
\subsection{Exact diagonalization results} \label{sec:exact}
\begin{figure}
  \includegraphics[width=\columnwidth]{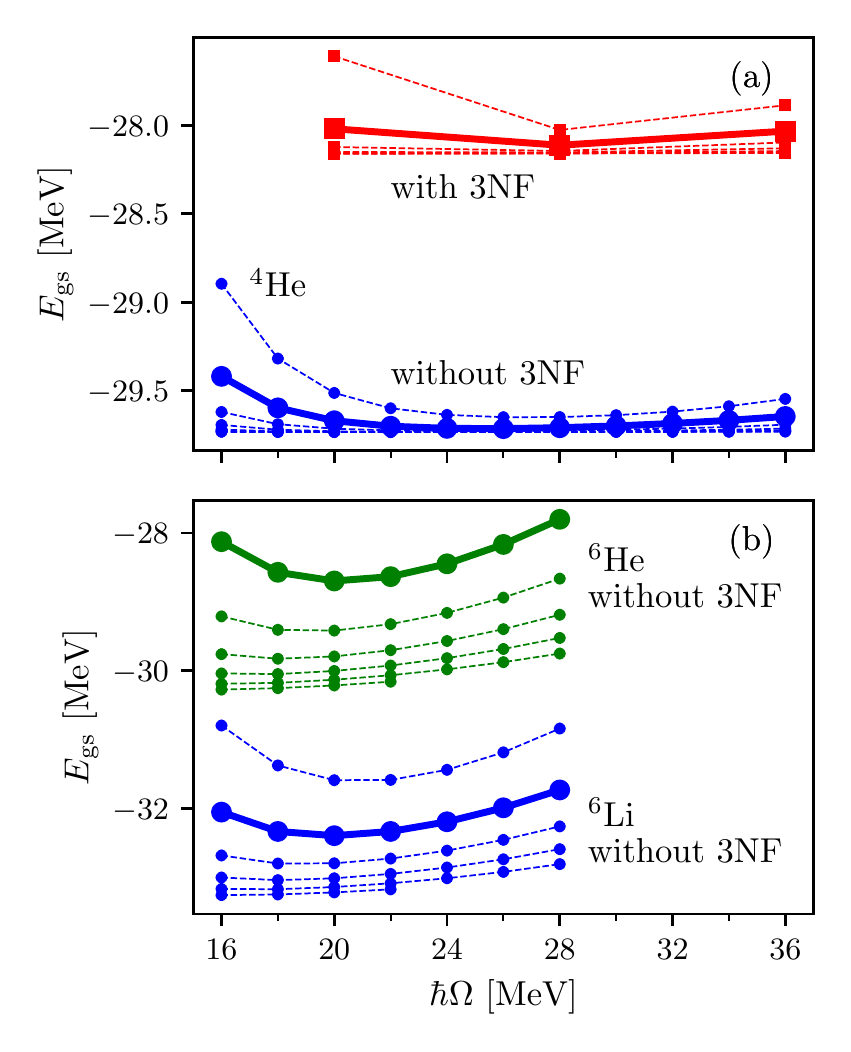}
  \caption{\NCSM{} convergence as a function of the basis frequency
    \hw{} for (a) \nuc{4}{He} and (b) \nuc{6}{Li}, \nuc{6}{He}
    computed with the \NNLO{} Hamiltonian without \threeNF{}. For
    \nuc{4}{He} we show also results with a fixed \threeNF{} ($\cD=-0.03$,
    $\cE=-0.20$). The lines connect
    results at fixed basis truncation; with variationally decreasing
    energies from $\Nmax=8$ to $\Nmax=20(18)$
    for $A=4(6)$. The thick solid lines correspond to $\Nmax=10(8)$
    which is the largest truncation used in the construction of
    emulators with \threeNF[s]{} for \nuc{4}{He} and \nuc{6}{Li}
    (\nuc{6}{He}). 
    These results show that the variational minimum appears near
    $\hw=20$~MeV for all isotopes. 
    \label{fig:convergence_hw}%
  }
\end{figure}

\begin{figure}
  \includegraphics[width=\columnwidth]{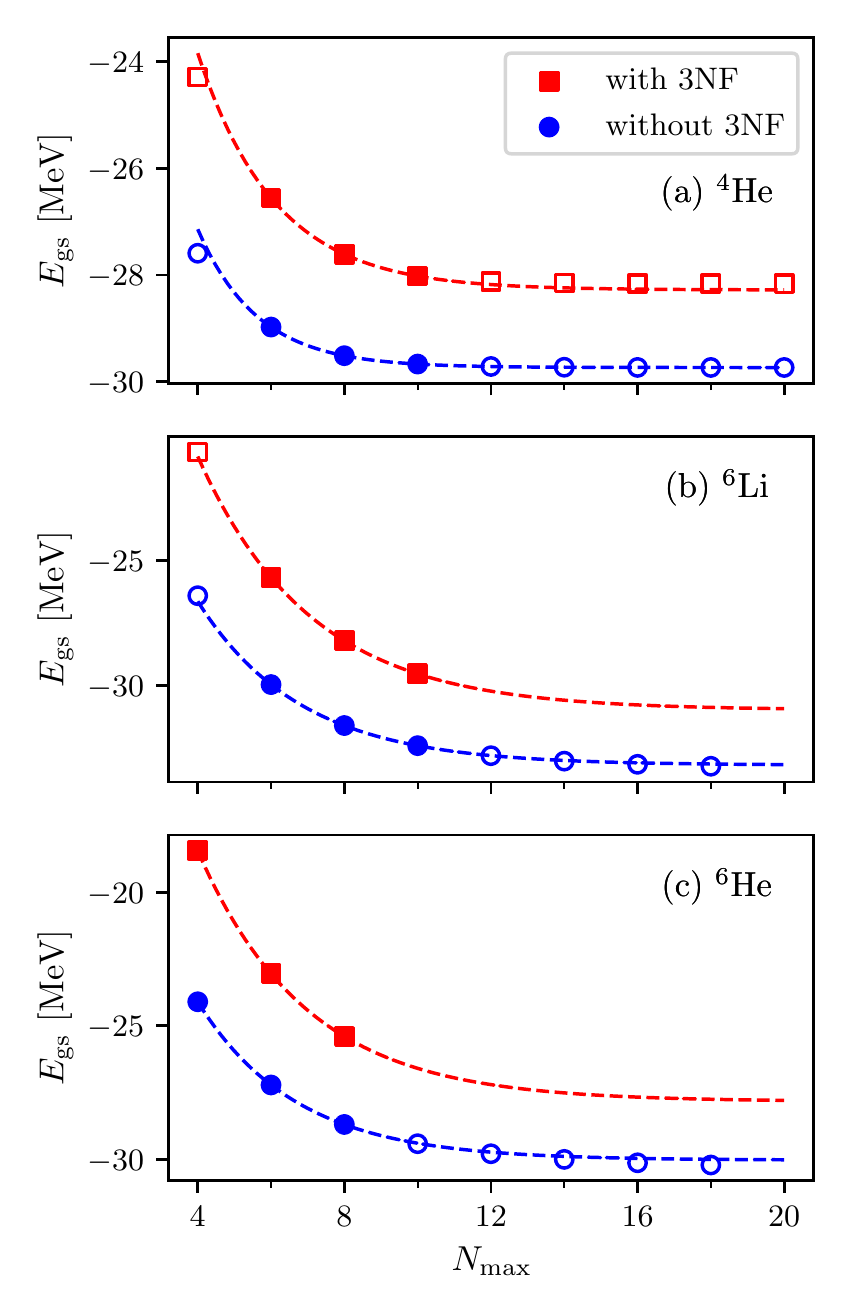} 
  \caption{\NCSM{} convergence as a function of the basis truncation
    \Nmax{} with a fixed basis frequency $\hw=20$~MeV. Results are
    shown for (a) \nuc{4}{He}, (b) \nuc{6}{Li}, and (c) \nuc{6}{He}
    computed with the \NNLO{} Hamiltonian with (square markers) and
    without (round markers) the \threeNF{}. Convergence is
    slower with the \threeNF{} as also indicated by the
    exponential extrapolation curves (dashed lines). The exponential
    fits are made with data from  
    $\Nmax=6,8,10$ for \nuc{4}{He} and \nuc{6}{Li}
    and 4,6,8 for \nuc{6}{He} as shown by the filled markers.
    \label{fig:convergence_Nmax}%
  }
\end{figure}

%
The \NCSM{} convergence for \nuc{4,6}{He} and \nuc{6}{Li} is shown in
Fig.~\ref{fig:convergence_hw} as a function of the basis frequency,
and in Fig.~\ref{fig:convergence_Nmax} as a function of basis
truncation \Nmax{} for fixed basis frequency $\hw=20$~MeV. 
In all calculations we use a fixed \NN{} force as described in
Sec.~\ref{sec:interaction}. For \nuc{4}{He} we also show results
including the \threeNF{} with a \MAP{} point estimate for the parameters $\cD=-0.03$
and $\cE = -0.20$ obtained  from \prA{} .

The thick solid lines in Fig.~\ref{fig:convergence_hw} correspond to
the largest truncation used with the \JNCSM{} code, namely
$\Nmax=10$ for \nuc{4}{He} and \nuc{6}{Li}, and $\Nmax=8$ for \nuc{6}{He}. Up to
this truncation we have been able to validate the ground-state energy results from
different \NCSM{} codes:
\nsopt{}~\cite{ekstrom2013} (for $A=4$),
\pantoine{}~\cite{forssen2017} (for $A=4,6$; with \twoNF{} only)
and \JNCSM{} (all nuclei; including full \threeNF{}).
In addition, results for $A=4,6$ nuclei using the $\NNLO_\mathrm{sat}$
interaction~\cite{ekstrom2015a} (with full \threeNF{}) have been
validated by finding $\approx 1$~keV differences in ground-state energies
from \JNCSM{} and the no-core
shell-model Slater determinant (\NCSD{}) code~\cite{NCSD}.

We conclude from these convergence studies that $\hw=20$~MeV is the optimal basis
frequency to use for this range of isotopes with these interactions. This basis 
    frequency will therefore be used for the construction of
    emulators. 

The fact that we use the \emph{ab initio} \NCSM{} implies that we make
controlled approximations when solving the many-body problem. This provides an
opportunity to quantify the magnitude of our method errors, 
which is a prerequisite when aiming for predictive power. We express the fully converged
prediction from the many-body problem as
\begin{equation}
  \obsMBinf(\lecs) = \obsMB(\lecs) + \errMB(\lecs),
  \label{eq:MB}
\end{equation}
where \obsMB{} is the result of our (basis truncated) \emph{ab initio} many-body solver
and \errMB{} is the corresponding
method error. In this work we will adopt a statistical model for this error
term. The convergence formula that forms the foundation for our error model is presented in the
following, while further details will be given in Sec.~\ref{sec:errors}.

In the case
of \HO{} basis expansion methods, such as the \NCSM{}, it has been shown that analyses in
terms of ultraviolet and infrared length scales offer much insight
into errors resulting from
finite basis truncations~\cite{coon2012,furnstahl2012,wendt2015,forssen2017}. In this
work, however, we restrict ourselves to a simpler analysis of the
convergence using an exponential form
\begin{equation}
  E(\Nmax) = E_\infty + a \exp(-b \Nmax),
  \label{eq:Nmax_extrapolation}
\end{equation}
which is fitted at a fixed frequency and with the parameter $E_\infty$
being an estimator of the converged result \obsMBinf{}. We note that the
convergence distance
\begin{equation}
  \Delta E_\infty(\Nmax) \equiv E_\infty -
  E(\Nmax)
  \label{eq:convdist}
\end{equation}
must be negative since the \NCSM{} is a variational method. We will
return to this extrapolation formula in Sec.~\ref{sec:errors} when we
introduce the statistical model for the method error.

The convergence of the ground-state energy as a function of the basis
truncation is shown in Fig.~\ref{fig:convergence_Nmax}. The validation
results at large \Nmax{} with \twoNF{}-only interactions indicate that the
exponential extrapolation slightly underestimates the convergence
distance for $A=6$ nuclei. We will later incorporate this finding into our
model of the method error. Furthermore, the observed convergence with respect to \Nmax{} with a full Hamiltonian
including \threeNF[s]{} is somewhat slower than the one with \twoNF{} only.

We have used the \LEC{} values from \textcite{Wesolowski:2021cni} to
study the order-by-order convergence of $A=4,6$ binding
energies. These point estimates correspond to a \MLE{} for the \LO{}
and \NLO{} Hamiltonians, and a \MAP{} for the \NNLO{} Hamiltonian. The
results from our \NCSM{} exact diagonalizations are shown in
Table~\ref{tab:orderbyorder}. Here we also show extrapolated results
with inclusion of the estimated mean method error, $\mu_{\delta E}$,
see Sec.~\ref{sec:errors}. We note in particular that our \LO{}
estimate corresponds to underbinding of these systems, which is in
stark contrast with many other studies that give strong overbinding at
\LO{}~\cite{carlsson2016,Maris:2020qne}. This difference can be traced
to the fit to low-energy scattering data
in~\cite{Wesolowski:2021cni} where the \chiEFT{}
expansion parameter $Q = m_\pi / \Lambda_b$ was employed in that domain---with
$m_\pi$ the pion mass and $\Lambda_b$ the \chiEFT{} breakdown
scale. As a result, the deuteron energy is $E_d = -0.6$~MeV at \LO{},
and this underbinding prevails for the $A=4,6$ systems.
\begin{table*}
   \begin{ruledtabular}
   \begin{tabular}{l|dd|dd|dd|c}
     &  \multicolumn{2}{c|}{LO$_{\text{MLE}}$} & \multicolumn{2}{c|}{NLO$_{\text{MLE}}$}
     &   \multicolumn{2}{c|}{N2LO$_{\text{MAP}}$} & \multicolumn{1}{c}{Experiment} \\
     & \multicolumn{1}{l}{$E_{\text{NCSM}}$} & \multicolumn{1}{l|}{$E _{\text{NCSM}} + \mu _{\delta E}$}
     & \multicolumn{1}{l}{$E_{\text{NCSM}}$} & \multicolumn{1}{l|}{$E _{\text{NCSM}} + \mu _{\delta E}$}
     & \multicolumn{1}{l}{$E_{\text{NCSM}}$} & \multicolumn{1}{l|}{$E _{\text{NCSM}} + \mu _{\delta E}$}
     \\
     \colrule  \bigstrut[t]
     $E(\nuc{4}{He})$ [MeV]
     & $-24.08$ & $-24.09$
     & $-30.21$ & $-30.21$
     & $-28.16$ & $-28.16$
     & $-28.296$~\cite{Tilley:1992zz}  \\
     $E(\nuc{6}{He})$ [MeV]
     & $-19.23$ & $-20.34$
     & $-28.49$ & $-28.79$
     & $-25.40$ & $-28.16$ 
     & $-29.271$~\cite{wang2012} \\
     $E(\nuc{6}{Li})$ [MeV]
     & $-19.83$ & $-21.32$
     & $-31.54$ & $-31.78$
     & $-29.52$ & $-31.13$ 
     & $-31.994$~\cite{wang2012} \\
   \end{tabular}
   \end{ruledtabular}
   \caption{Results at \LO{}, \NLO{}, and \NNLO{} for the $A=4,6$
     binding energies. For \LO{} and \NLO{}, these correspond to the
     \MLE{} from~\cite{Wesolowski:2021cni}. The \NNLO{} computations
     use \piN{} and \NN{} LECs
     from~\cite{siemens2017,Wesolowski:2021cni} and the \MAP{} point
     estimate is obtained with $\cD=-0.03$ and $\cE=-0.20$.
     The variational minimum obtained in the largest
     model space is shown in the $E_{\text{NCSM}}$ column for each
     order, while the extrapolated result~\eqref{eq:Nmax_extrapolation} with error
     estimates are shown in the $E _{\text{NCSM}} + \mu _{\delta E}$
     column, where $\mu _{\delta E} \equiv \mathbb{E} \left[ \errMB{} \right]$.
     Experimental data are from~\cite{Tilley:1992zz,wang2012} and
     have uncertainties that are negligible in this context.
   \label{tab:orderbyorder}    %
 }
\end{table*}
%
\subsection{Eigenvector continuation emulators} \label{sec:emulator}
%
The specific aim in this work is the computation of a \PPD{} for the
ground-state energies of several many-body systems. This requires the
ability to solve the \MBSE{} repeatedly for different nuclear systems
and for many samples from the Hamiltonian parameter \PDF{}. In this
work we will solve for \nuc{4}{He}, \nuc{6}{Li}, and \nuc{6}{He} with
$\sim 2.5 \cdot 10^6$ samples from \prA{}. This is achieved by
exploiting the method of \EC{}~\cite{frame2017,frame2019,konig2019} to
mimic the solution of the full problem with high accuracy at a
fraction of the computational cost.

Given a Hermitian matrix that depends smoothly on some continuous
parameters---in our case \(H(\cD,\cE)\)---this method can be used to
construct an emulator that performs very well in a large parameter
domain. The training of such an emulator requires a small set of
training vectors obtained by solving the full problem for a
corresponding set of training points in (\cD,\cE) space.

For each nucleus (\nuc{4,6}{He}, \nuc{6}{Li}) and model space ($\Nmax
\in \{4,6,8,10\}$), we compute the ground state by
solving Eq.~\eqref{eq:mbse} using \JNCSM{} for a total of \(16\)
training points $\{{\cDk},{\cEk}\}_{k=1}^{16}$.
Eight of these training points correspond to random draws from the posterior \prA{}
that is shown in
Fig.~\ref{fig:cDcE_pdf}(a), while the remaining eight were drawn from
a large square: $\cD,\cE \sim \mathcal{U}(-2.5,2.5)$. After performing
diagonalizations---using the eigenvector convergence
criterion with $\epsilon_2=10^{-6}$---this results in 16 training vectors for each nucleus and
model space, i.e.
\begin{equation}
  \ket{\Phi_k} = \ket{\Psi({\cDk},{\cEk})}, \,\,\, k=1,2,\ldots,16
\end{equation}

When using \EC{}, the Hamiltonian is projected onto the subspace that
is spanned by the training vectors. We denote our subspace-projected
Hamiltonian as \(M(\cD,\cE)\) and its matrix elements are computed as
\begin{equation}
  \begin{aligned}
  \left(M(\cD,\cE)\right)_{i,j} =& \mel{\Phi_i}{H_0}{\Phi_j} \\
  +&\cD{}\mel{\Phi_i}{H_1}{\Phi_j} \\
  +&\cE{}\mel{\Phi_i}{H_2}{\Phi_j} \\
  =& \left(M_0\right)_{i,j}+\cD \left(M_1\right)_{i,j} + \cE \left(M_2\right)_{i,j},
  \end{aligned}
\end{equation}
where we note that \(M_0\) \(M_1\) and \(M_2\) are $16 \times 16$
matrices that can be computed once
per nucleus and model space.

There is no guarantee that the training vectors are orthogonal.
Therefore, we also construct the norm matrix \(N\) such that
\begin{equation}
  \left(N\right)_{i,j} = \braket{\Phi_i}{\Phi_j}.
\end{equation}
The subspace-projected \MBSE{} becomes a generalized eigenvalue problem
\begin{equation} \label{eq:generalized_eigenvalue_problem}
  M(\cD,\cE)\vec{v}(\cD,\cE)=\lambda(\cD,\cE) N\vec{v}(\cD,\cE),
\end{equation}
where \(\lambda\) (\(\vec{v}\)) is the eigenvalue (eigenvector).
Since we are projecting the \NCSM{} Hamiltonian onto a subspace, the \EC{} method
is variational and therefore
\(E_{\rm{gs}}^{\rm{EC}} = \min{\{\lambda(\cD,\cE)\}} \geq
E_{\rm{gs}}(\cD,\cE)\). It turns out that $E_{\rm{gs}}^{\rm{EC}}$ is a
very good approximation of the NCSM ground-state energy.  

\begin{figure}
 \includegraphics[width=\columnwidth]{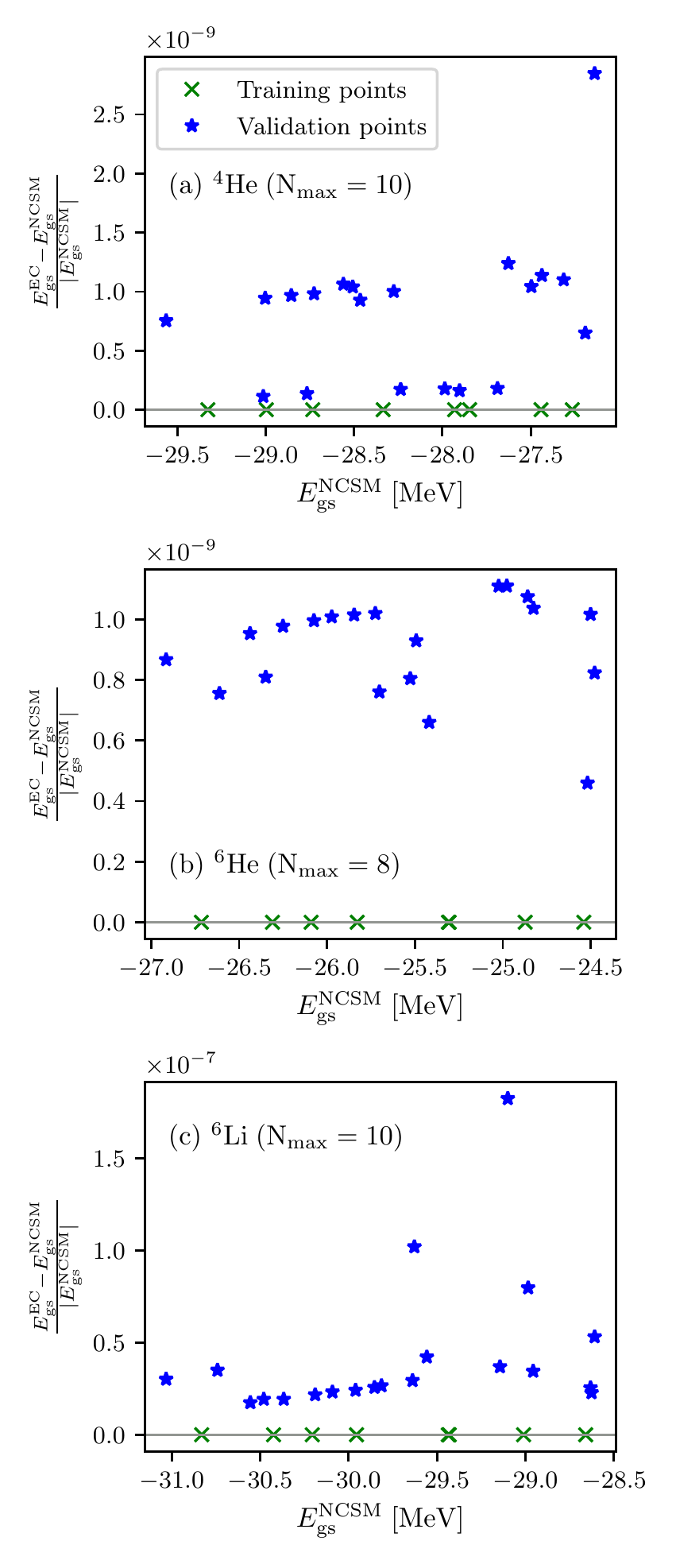} 
  \caption{Cross-validation of the emulators by comparing the relative
    differences between the full solution of the Schr\"odinger
    equation (at fixed basis space truncation \Nmax{}) and those of
    the \EC{} emulators at 20 validation points (blue stars). We also demonstrate the
    validation at the training points (green crosses), where the
    subspace eigensolution should be identical to the full one. Note
    that the ground state energies from the emulator are guaranteed to
    be an upper bound on the exact NCSM energies.
    \label{fig:CV}%
  }
\end{figure}

A validation of the \EC{} emulators is performed by selecting an
additional 40 (\cD{},\cE{})-samples; \(20\) of which are new draws
from the posterior \prA{}, and \(20\) from the large square: $\cD,\cE \sim \mathcal{U}(-2.5,2.5)$.
We then compare emulated ground-state energies
$E_{\rm{gs}}^{\rm{EC}}$, obtained by solving
Eq.~\eqref{eq:generalized_eigenvalue_problem}, and full numerical
solutions $E_{\rm{gs}}$ obtained with \JNCSM{} using an eigenvalue
convergence criterion with $\epsilon_1=10^{-7}$. 
The results of this validation are shown in Fig.~\ref{fig:CV} for the
largest \Nmax{} used for each nucleus. Here we focus on the most
interesting region in parameter space and include the 20 validation points (blue stars) and eight training points
(green crosses) that are drawn from the posterior distribution (see Fig.~\ref{fig:cDcE_pdf}). 

As can be seen, the relative difference of the validation points are
very small for all three nuclei. For the two helium isotopes this
difference is around \(10^{-9}\), while for \nuc{6}{Li} it is around
\(10^{-7}\), which is close to the convergence criterion that we used
for the Lanczos algorithm in \JNCSM{}. Therefore, we conclude that
\EC{}-emulated ground-state energies are basically as accurate as the
same values computed in the full Hilbert space using \JNCSM{} for
\(\cD{}\), \(\cE{}\) values within the posterior \prA{}. In fact, we
find almost the same precision in the large square of (\cD,\cE)
parameter values, except for points that produce extreme energies
($E_{\rm{gs}}< -100$~MeV) for which the relative difference becomes
$\sim 10^{-2}$.
%
%
\section{Posterior predictive distributions} \label{sec:ppd}
%
In this section we will compute and study the NCSM \PPD{} which we
define as
\begin{equation} 
    \text{PPD}_{\text{NCSM}} = \{\obsMBset(\lecs) : \lecs \sim \pr(\lecs
  \given \data, I) \}.
  \label{eq:trunc_model_ppd}
\end{equation}
This \PPD{} is the set of all model predictions computed over likely
values of the \LEC[s]{}, i.e., drawing from the posterior \PDF{} for the
$\lecs$. Such results are shown in
Figs.~\ref{fig:trunc_model_marginal_ppd} and
\ref{fig:trunc_model_bivariate_ppd} and discussed below. They
illustrate the parametric uncertainty. Note, however, that this
NCSM
\PPD{} does not include the uncertainties that stem from the
truncation of our many-body solver (method uncertainty) or the
truncation of our EFT (model uncertainty).

Such additional sources of uncertainty are rather consolidated in
the full \PPD{}, where we incorporate our knowledge of the model discrepancy, method uncertainty, and
emulator error terms. The full \PPD{} is then defined, in analogy with
Eq.~\eqref{eq:trunc_model_ppd},  as the set evaluation of
$\obsset{}(\lecs)$ which is the sum
\begin{equation}
  \obsset(\lecs) = \obsMBset(\lecs) + \errMBset(\lecs) + \errEFTset + 
  \erremset. 
  \label{eq:fullstatmodel}
\end{equation}
We will consider each of the error terms \errset[i]{} in Eq.~\eqref{eq:fullstatmodel} as random variables with
distributions that we will discuss in detail in Sec.~\ref{sec:errors}.
It should be noted that we explicitly assume that the method error
might depend on the \LEC{}s and that the emulator error is negligible
compared to the other uncertainties, as shown in
Sec.~\ref{sec:emulator}, such that the last term will be disregarded in the following.
The final predictions, considering relevant uncertainties, are
discussed in Sec.~\ref{sec:errors} and displayed in
Figs.~\ref{fig:model_ppd} and \ref{fig:model_ppd_levels}.
%
\subsection{NCSM posterior predictive distribution} \label{sec:model_ppd}
%
The relevant question on the magnitude of parametric uncertainties in
predictions made with our nuclear \emph{ab initio} approach can be addressed with the NCSM
\PPD~\eqref{eq:trunc_model_ppd}. For this purpose we utilize the
few-body-constrained \LEC{}
posterior~\cite{Wesolowski:2021cni} discussed in Sec.~\ref{sec:interaction}, together with our \EC{}
emulators for $A=4,6$ binding energies (see Sec.~\ref{sec:emulator}). These fast and accurate emulators
make it possible for us to predict the \nuc{4,6}{He} and
\nuc{6}{Li} binding energies for $\sim 2.5 \cdot 10^6$ samples 
from \prA{}~\eqref{eq:prA}.

\begin{figure}
  \includegraphics[width=0.95\columnwidth]{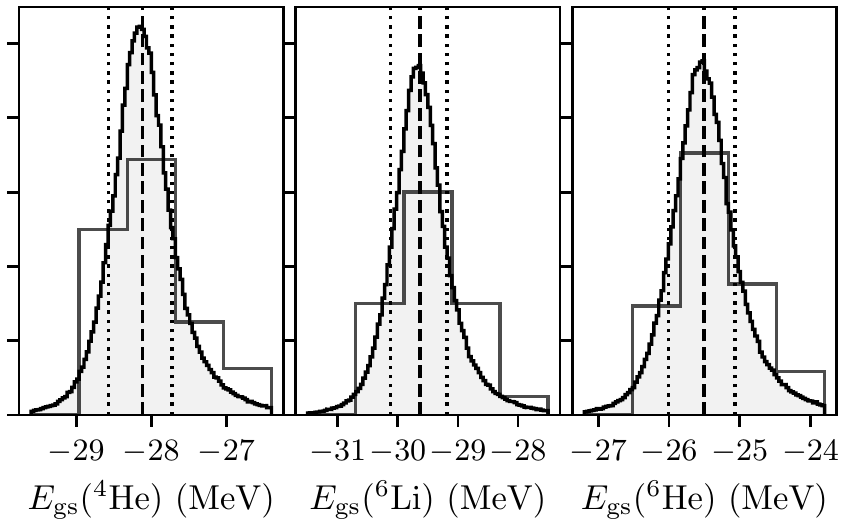} 
  \caption{Marginal distributions from
    \PPD{}$_\text{NCSM}$~\eqref{eq:trunc_model_ppd} illustrating the
    parametric uncertainties
    from the posterior distribution for $(\cD{}, \cE{})$~\eqref{eq:prA}
    propagated to ground-state energies of \nuc{4}{He}, \nuc{6}{Li} 
    computed with $\Nmax{}=10$ and \nuc{6}{He} with $\Nmax{}=8$. The
    median value and the
    68\% Bayesian credible interval is indicated by dashed and dotted
    lines, respectively. See also
   Table~\ref{tab:ppd}. The open, grey
    histograms on the diagonal represent low-statistics results with
    only 25 samples (five bins).
    \label{fig:trunc_model_marginal_ppd}%
  }
\end{figure}

The resulting marginal distributions are shown in Fig.~\ref{fig:trunc_model_marginal_ppd} together with the medians and the 68\%
Bayesian credible intervals (obtained as the smallest region that
contains 68\% of the probability mass). Numerical values for these
summary statistics are presented in Table~\ref{tab:ppd}.

The shape of the \nuc{4}{He} distribution can be compared with the
results shown in the diagonal panel (second row) of Fig.~3 in
Ref.~\cite{Wesolowski:2021cni}. However, it should be noted that while our
\nuc{4}{He} emulator was constructed with $\Nmax=10$ in the $M$-scheme
\NCSM{}, those of \textcite{Wesolowski:2021cni} were based on a
Jacobi-coordinate \NCSM{} implementation and used $\Nmax=18$. This
gap in model space results in $\sim 100$~keV energy difference.

Accurate emulators might not always be available for the observable of
interest. For that reason we demonstrate a low-statistics
representation of the same results by open histograms in
Fig.~\ref{fig:trunc_model_marginal_ppd}. These results
are based on merely 25 samples from \prA{}. Both the median and the standard
deviation of the predictions made with the low-statistics
representation agrees to within 10~keV with the summary statistics
obtained from the full list of samples. Although small statistics might not allow
a very precise quantification of credible intervals, it is clear
that valuable parametric uncertainty estimates can still be extracted.

\begin{figure}
  \includegraphics[width=0.95\columnwidth]{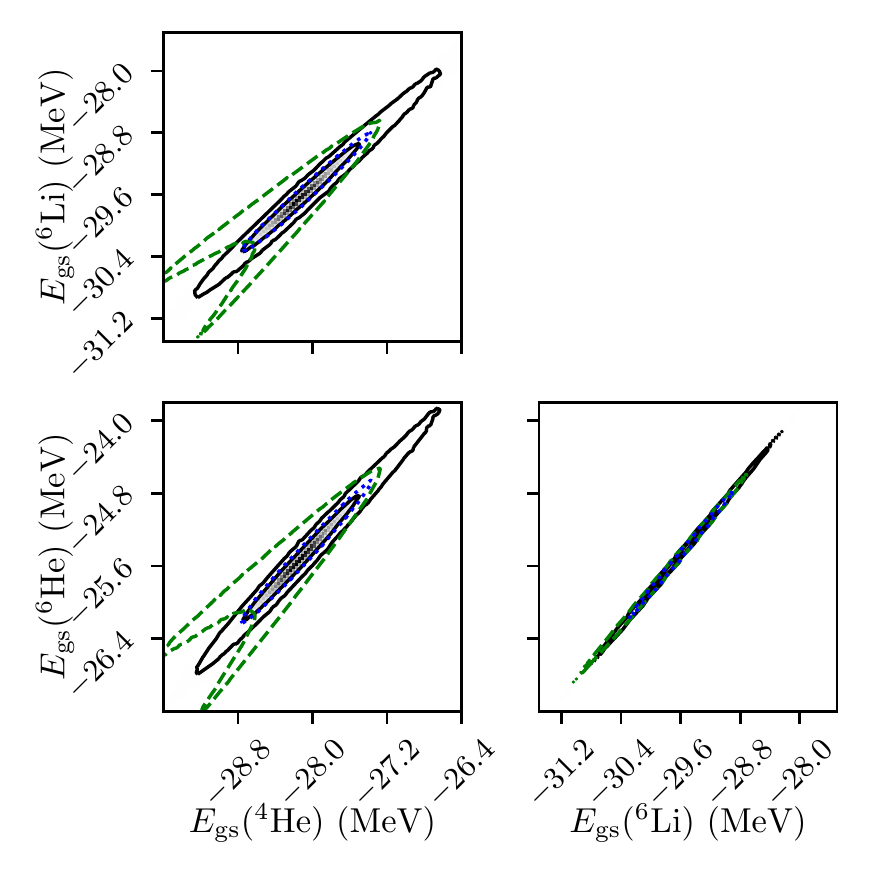} 
  \caption{Bivariate distributions from
    \PPD{}$_\text{NCSM}$~\eqref{eq:trunc_model_ppd} illustrating the
    parametric uncertainties
    from the posterior distribution \prA{}~\eqref{eq:prA}
    propagated to ground-state energies of \nuc{4}{He}, \nuc{6}{Li} 
    computed with $\Nmax{}=10$ and \nuc{6}{He} with $\Nmax{}=8$. The
    68\% and 95\% probability mass regions are shown by the
    level curves. For comparison, the
    95\% regions for the posteriors \prB{} and \prC{} are shown with
    green dashed and blue dotted curves, respectively. 
    \label{fig:trunc_model_bivariate_ppd}%
  }
\end{figure}

The bivariate distributions from \PPD{}$_\text{NCSM}$ are shown in
Fig.~\ref{fig:trunc_model_bivariate_ppd}. We observe very strong
correlations between the predicted binding energies, as could be
expected. This means that the uncertainties in relative energies will
be much smaller, as we will discuss in more detail below. For
comparison we also show the bivariate distributions that are obtained
when using samples from the alternative parameter posteriors \prC{}
and \prB{}, see Fig.~\ref{fig:cDcE_pdf} and the discussion in
Sec.~\ref{sec:interaction}. The fact that the truncation-error
parameters \cbar{} and $Q$ were fixed leads to a much narrower (more
Gaussian) distribution of $(\cD,\cE)$ in \prC{}, which is also
reflected in the tighter NCSM \PPD{} (blue, dotted region).

Even more interesting is the behavior of the NCSM \PPD{} from
\prB{}. As shown in Fig.~\ref{fig:cDcE_pdf}(c), the use of highly
correlated calibration observables, $E(\nuc{3}{H})$ and
$E(\nuc{4}{He})$, gives an effective parameter constraint in a single
direction which leads to a very broad range of allowed
$(\cD,\cE)$-values. The propagation of that parametric uncertainty to predictions
of the $A=6$ binding energies reveals if those new observables would be
able to provide additional, independent constraints. We actually find
that the bulk region of the NCSM \PPD{} obtained from \prB{} does
largely overlap with the ones from \prA{} and \prC{} that are informed
by a larger set of few-body observables. This indicates that their
propagated predictions for $A=6$ binding energies are very similar. However, the most extreme samples
from \prB{}, i.e., $\cD \sim -5$ and $\cD \sim +5$, correspond
respectively to the upper and lower ``legs'' of the green, dashed 95\%
confidence regions shown in the left column of
Fig.~\ref{fig:trunc_model_bivariate_ppd}.
\begin{table}
      \begin{ruledtabular}
      \begin{tabular}{l|dc|dc}
        Observable &  \multicolumn{2}{c|}{NCSM \PPD{}} & \multicolumn{2}{c}{Full \PPD{}}\\
        & \multicolumn{1}{c}{median} & CI (68/95\%)
        & \multicolumn{1}{c}{median} & CI (68/95\%) \\
        \colrule  \bigstrut[t]
        $E(\nuc{4}{He})$ & -28.11 &  [-0.46,+0.39] & -28.23                             &  [-0.80,+0.75]  \\ 
                   &      &   [-0.99,+1.20] &                                           &   [-1.59,+1.65] \\ 
         $E(\nuc{6}{Li})$ & -29.63 &  [-0.50,+0.44] & -31.34                              &  [-0.92,+0.89]  \\ 
             &      &   [-1.09,+1.35] &                                                 &   [-1.84,+1.92] \\ 
         $E(\nuc{6}{He})$ & -25.51 &  [-0.50,+0.44] & -28.42                            &  [-0.96,+0.95]  \\ 
             &      &   [-1.09,+1.35] &                                                 &   [-1.97,+2.00] \\ 
          $S_d(\nuc{6}{Li})$ & -0.71 &  [-0.08,+0.08] &  0.89                                  &  [-0.44,+0.44]  \\ 
             &      &   [-0.21,+0.21] &                                                 &   [-0.87,+0.88] \\ 
         $S_{2n}(\nuc{6}{He})$ & -2.61 &  [-0.08,+0.07] &  0.19                                 &  [-0.61,+0.58]  \\ 
             &      &   [-0.20,+0.20] &                                                 &   [-1.17,+1.19] \\ 
       $Q_{\beta^-}(\nuc{6}{He})$ &  4.90 &  [-0.01,+0.00] &  3.71                                 &  [-0.65,+0.64]  \\ 
             &      &   [-0.01,+0.02] &                                                 &   [-1.26,+1.27] \\ 
      \end{tabular}
  \end{ruledtabular}
  \caption{Median predictions and 68\% (95\%) Bayesian credible
    intervals shown in the first (second) row for each observable. All
    values are in MeV and the intervals are presented as differences
    from the median. The NCSM \PPD{} is obtained by using the raw
    output from our emulators (\nuc{4}{He}, \nuc{6}{Li} 
    computed with $\Nmax{}=10$ and \nuc{6}{He} with $\Nmax{}=8$). The
    full \PPD{}, however, includes also statistical models for the method and
    model errors (see text for details). Experimental energies
    are listed in Table~\ref{tab:orderbyorder} and result in the energy
    differences: $S_d(\nuc{6}{Li}) = 1.473$~MeV, $S_{2n}(\nuc{6}{He})=0.975$~MeV, and
    $Q_{\beta^-}(\nuc{6}{He})=3.505$~MeV. 
  \label{tab:ppd}
}
\end{table}
\subsection{Full posterior predictive distribution} \label{sec:errors}
%
The NCSM \PPD{} does not reflect uncertainties associated
with the truncation of the many-body model space, nor with the
truncation of the \EFT{} expansion. 
Let us therefore incorporate models for these additional, relevant errors
into our final predictions. We start from Eq.~\eqref{eq:fullstatmodel} that
sums up the assumed link between model and reality. As previously mentioned
we will neglect emulator errors ($\errem \approx 0$) and we will
assume that the EFT error is independent on the \LEC{} parameters such that
we have
\begin{equation}
  \obsset(\lecs) \approx \obsMBset(\lecs) + \errMBset(\lecs) + \errEFTset.
  \label{eq:statmodel}
\end{equation}
We will now discuss our statistical models for the two final terms in
this expression.

\subsubsection{Method errors} \label{sec:methoderrors}
%
 The method error is estimated from the observed convergence behavior
and our previous experience with the \NCSM{} using also other interaction
models. In particular, we know that the \NCSM{} is a variational
approach such that results obtained at truncated model spaces represent upper
bounds. This implies that the expectation value for the method error
of the total energy observable $i$ is negative
\begin{equation}
  \mathbb{E} \left[ \errMB{}_{,i}(\lecs) \right] \equiv \mu_{\delta E,i}(\lecs) < 0.
\end{equation}
Since we have constructed emulators for different model spaces, $\Nmax
\in \{ 4,6,8,10 \}$, we can perform the
extrapolation~\eqref{eq:Nmax_extrapolation} for each set of \LEC[s]{}
and extract an \lecs-dependent convergence distance $\Delta
E_{\infty,i}(\lecs)$ in Eq.~\eqref{eq:convdist}. However, the
convergence tests in Sec.~\ref{sec:exact} together with previous
experience~\cite{forssen2008,wendt2015,forssen2017} tells us that this
simple extrapolation tends to underestimate the convergence distance
by a few ten percent. To incorporate this knowledge we estimate the
mean value of the method error as
\begin{equation}
  \begin{split}
  \mu_{\delta E,i}(\lecs) &= \Delta E_{\infty,i}(\lecs) +
  \sigma_{\text{NCSM},i}, \\
  \text{with } \sigma_{\text{NCSM},i}(\lecs) &= 0.2 \Delta E_{\infty,i}(\lecs).
  \end{split}
  \label{eq:methoderr_mean}
\end{equation}
The mean value for $\Delta E_{\infty}$ for the \LEC{} samples from
\prA{} is found to be $-1.46$ ($-2.46$)~MeV for \nuc{6}{Li}
(\nuc{6}{He}). The standard deviation for $\Delta E_{\infty}$ is
very small at $50 (120)$~keV and we therefore simplify the assignment by making
$\lecs$-independent estimates: $\sigma_{\text{NCSM}} =  -290$
($-490$)~keV for \nuc{6}{Li}
(\nuc{6}{He}). Furthermore, the square of this correction is used as the variance of the assigned
method error.

For \nuc{4}{He} we find that the mean value for $\Delta E_{\infty}$ is
$-0.26$~MeV with a very small standard deviation of $23$~keV. For this
short extrapolation distance we find that the exponential form actually
overestimates the missing binding energy and gives an $E_\infty$ that is about
120~keV below the converged result, see
Fig.~\ref{fig:convergence_Nmax}(a). Consequently, we assign
$\sigma_{\text{NCSM}} =  +120$~keV for this nucleus.

In summary, we model the distribution of the method error random
variables $\errMB{}_{,i}(\lecs)$ as normal distributions
\begin{equation}
  \errMB{}_{,i}(\lecs) \sim \mathcal{N}
  \left(\mu_{\delta E,i}(\lecs), \sigma_{\text{NCSM},i}^2 \right),
  \label{eq:methoderr}
\end{equation}
with an \lecs{}-dependence in the mean value---stemming from the observed
\Nmax{}-convergence---and a nucleus-dependent convergence
uncertainty that both corrects the mean and defines the estimated variance.
\subsubsection{Model errors} \label{sec:modelerrors}
\begin{figure}
  \includegraphics[width=\columnwidth]{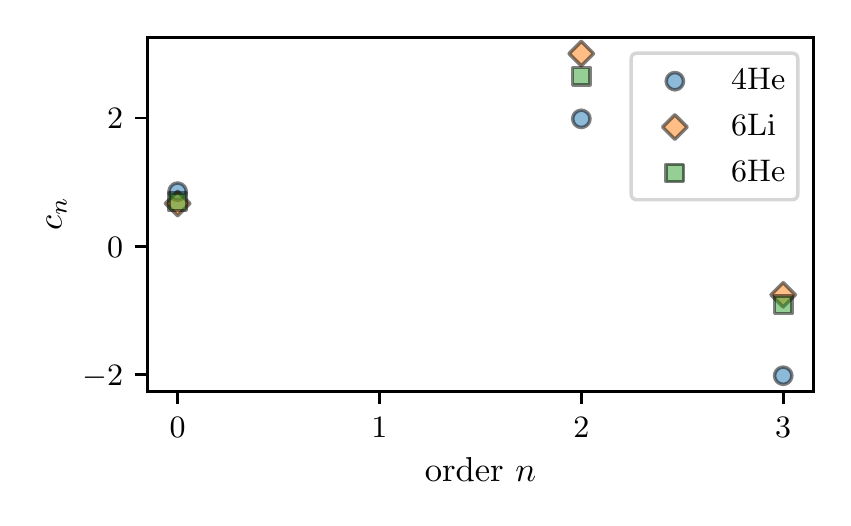}
 \caption{Observable coefficients, $c_n$, for $A=4,6$ binding energies
   up to \NNLO{} obtained using a chiral expansion parameter $Q=1/3$
   for all three observables.
  \label{fig:observablecoeff}%
}
\end{figure}

%
We create a statistical model for the EFT model discrepancy
\errEFTset{} based on the observed terms in the assumed EFT
convergence pattern~\eqref{eq:obsSum}. Point estimates for predictions
at each order from Table~\ref{tab:orderbyorder} are converted to
observable coefficients assuming a fixed value for the expansion
parameter, $Q=0.33$ from Ref.~\cite{Wesolowski:2021cni}. The resulting
coefficients are shown in
Fig.~\ref{fig:observablecoeff}. It is obvious that the corrections to the three binding energies are similar at
each order which indicates that the
observable coefficients are strongly correlated. 
We will assume that these coefficients are \iid{} random variables drawn from a single multivariate
normal distribution: $\obscset[n] \sim \mathcal{N}\left( 0, \cov \right)$, with the
covariance matrix expressed in terms of its
diagonal elements and a simple correlation matrix
\begin{equation}
  \begin{split}
    \cov &= \stddev \corr \stddev, \\
    \text{where } \stddev{}_{ij} &= \left\{
      \begin{array}{ll} \cbar & \text{for } i=j , \\
        0 &  \text{for }  i\neq j ,
      \end{array}
    \right. \\
    \text{and } \corr{}_{ij} &= \left\{
      \begin{array}{ll} 1 & \text{for } i=j , \\
        \rho &  \text{for }  i\neq j .
      \end{array}
    \right.
    \end{split}
  \label{eq:covariance}
\end{equation}
A straightforward \MLE{}, with the likelihood based on the data shown in Fig.~\ref{fig:observablecoeff},
gives $\cbar=1.7$ and $\rho=0.9$. We will use those as fixed
parameters in the following.

For this given model of the EFT observable coefficients, all neglected
terms beyond $\mathcal{O}(Q^k)$ in Eq.~\eqref{eq:obsSum} can be summed
to give a distribution for the model error
\errEFTset{}~\cite{Furnstahl:2015rha}. Specifically, with $\obscset[n] \sim \mathcal{N}\left( 0, \cov \right)$ 
this sum can be performed analytically~\cite{Wesolowski:2018lzj,Melendez:2019izc} and we find 
\begin{equation}
\begin{split}
  \errEFTset &\sim \mathcal{N} \left( 0, \cov[th] \right), \\
  \text{with } \cov[th]{}_{,ij} &= \cbar^2 \obsref{}_{,i} \obsref{}_{,j} \frac{Q_i^{k+1}Q_j^{k+1}}{1-Q^2} \corr{}_{ij}.
  \end{split}
  \label{eq:modelerr}
\end{equation}
Note that we use the same $Q_i=0.33$ for all $A=4,6$ observables.
\subsubsection{Full sampling of the \PPD{}}
\begin{figure*}
  \includegraphics[width=\textwidth]{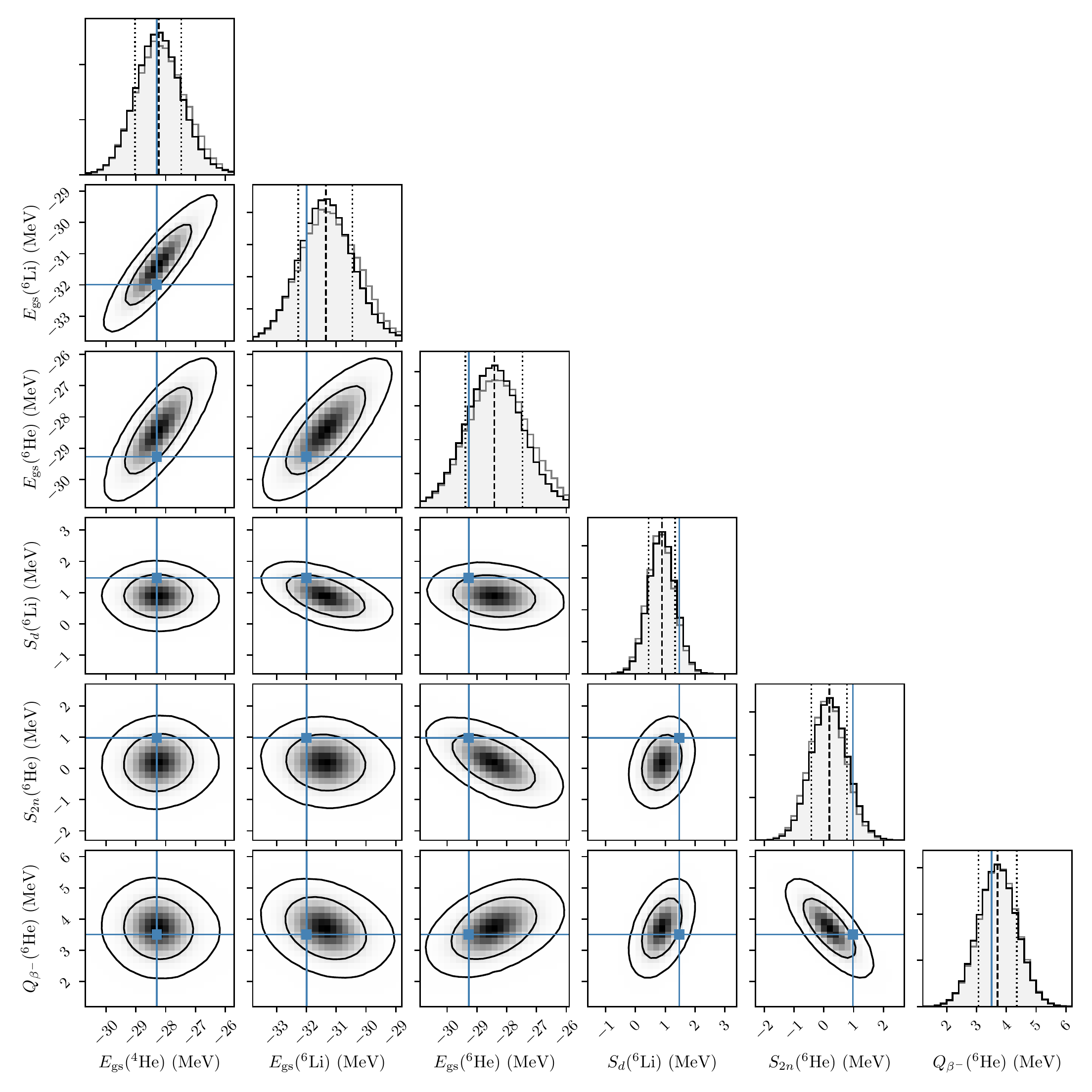}
 \caption{Full \PPD{} for binding energies and thresholds including both method
   and model (EFT truncation) uncertainties. The dashed (dotted), vertical lines on the
   diagonal show the median (68\% credible interval), while the
   blue, solid lines indicate the experimental values. See also
   Table~\ref{tab:ppd}. The open, grey
    histograms on the diagonal represent low-statistics results based on
    only 25 \LEC{} samples (see text for details). The level
   curves in the off-diagonal panels show the
    68\% and 95\% probability mass regions of the bivariate
    distributions. 
   \label{fig:model_ppd}%
 }
\end{figure*}

%
Results from sampling of the full \PPD{} are shown in
Fig.~\ref{fig:model_ppd}.
In practice, we construct this full \PPD{} with $\sim 2.5 \cdot 10^6$
samples from the NCSM
\PPD{} with the addition of the random variables \errMBset{} and
\errEFTset{} sampled from the distributions \eqref{eq:methoderr} and
\eqref{eq:modelerr}, respectively. The sampling from the error
distributions is simplified by the fact that the only
\LEC{}-dependence sits in the mean value of the method error via the
$\Delta E_\infty(\lecs)$ term in Eq.~\eqref{eq:methoderr_mean}. We are able
to compute this extrapolation term for any set of $\lecs$ since we
have created accurate emulators for $\Nmax \in \{4,6,8,10\}$.

The separation energies, $S_d$ for
\nuc{6}{Li} and $S_{2n}$ for \nuc{6}{He}, and the
$\nuc{6}{He}_\mathrm{gs} \to \nuc{6}{Li}_\mathrm{gs}$ beta-decay
$Q$-value are also included. Predictions for these relative
observables will now make sense since we have incorporated both method
errors and correlated model discrepancies. The former contains
information on the convergence of the many-body solver, while the
latter embodies the EFT truncation error and makes sure that we do not
overestimate its effect when propagated to differences of correlated
observables.
The final predictions for $A=6$ total energies, shown in the second
and third column of Fig.~\ref{fig:model_ppd}, provides hints for
systematic underbinding. We speculate that there is a possibility that
nuclear interactions that are constructed using more relaxed
low-energy constraints (causing the deuteron energy to be less
accurately reproduced) can lead to systematic underbinding in larger
systems. However, the statistical evidence for this proposition is not very
significant.

We also employed a low-statistics set of samples with the resulting full
\PPD{} shown by the open, grey histograms in the diagonal panels of
Fig.~\ref{fig:model_ppd}. Here we start from just 25 samples from
\prA{} and the corresponding NCSM predictions. We then resample
25,000 times from these 25
predictions and add samples from the method and model
errors. The close resemblance of the resulting marginal \PPD{}
distributions with the high-statistics version shows that in this case
it is indeed
possible to propagate errors and extract relevant uncertainty
quantification estimates also with a relatively small number of model
predictions. 

\begin{figure}
  \includegraphics[width=\columnwidth]{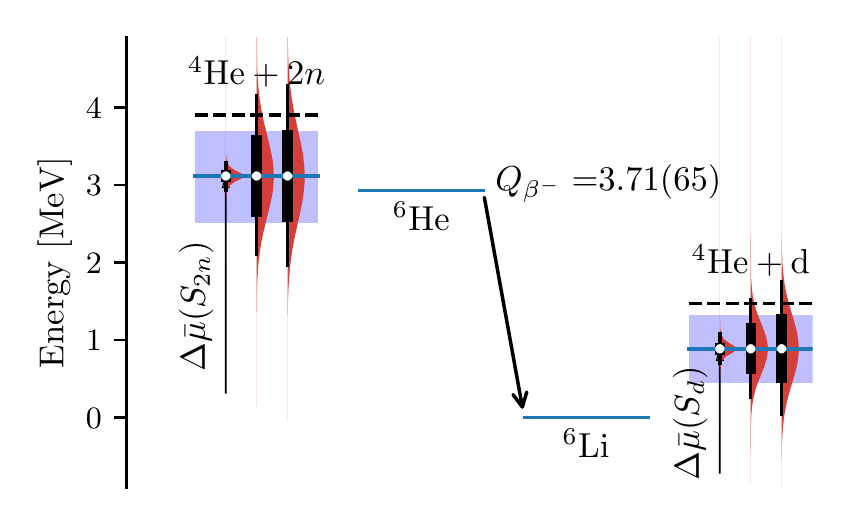}
 \caption{$A=6$ level
   scheme. Dashed lines show experimental thresholds for
   $\nuc{4}{He}+2n$($\nuc{4}{He}+\mathrm{d}$) relative
   \nuc{6}{He}(\nuc{6}{Li}) while the blue line and band show the
   median and 68\% credible interval from the full \PPD{}. The red distributions,
   from left to right, show the evolution of the \PPD{} as we go from the
   NCSM prediction, \PPD{}$_\text{NCSM}$, to the inclusion of method errors, and
   finally including the EFT truncation error---with thick (thin)
   vertical lines indicating the 68\%(95\%) credible interval. Note that the
   NCSM prediction for each threshold has been shifted by the mean values of
   the relevant method errors. The
   uncertainty in the $\beta^-$-decay $Q$-value is dominated by the
   method (\Nmax{}-extrapolation) uncertainty.
  \label{fig:model_ppd_levels}%
}
\end{figure}

Finally, we show the $A=6$ level scheme in
Fig.~\ref{fig:model_ppd_levels} with relevant uncertainties. This result
demonstrates the precision that can be expected in our \emph{ab
  initio} approach. The sequences of \PPD{} distributions show that it
is in fact the uncertainty in the many-body solver, and the process of
extrapolating to infinite model space, that is responsible for the
main fraction of the total error budget. This situation might be
different for other many-body systems and using other many-body
methods. 
We expect that already an extension to $\Nmax=10, 12$ for
\nuc{6}{He} and $\Nmax=12$ for \nuc{6}{Li} would significantly reduce
these method errors. 
%

\section{Summary and outlook} \label{sec:summary}
%
The main findings and conclusions of this study are:
\begin{itemize}
\item{\textbf{This work introduces \JNCSM}
    ---a new $M$-scheme NCSM code that uses on-the-fly Hamiltonian
    matrix construction in a \SD{} basis with full inclusion of
    \threeNF[s]{}. This code performs computationally efficient
    diagonalization using the Lanczos algorithm and can employ various
    convergence criteria. In particular, we perform single-node
    computations up to $\Nmax=10$ for \nuc{6}{Li} focusing on
    obtaining well-converged eigenvectors.
  }
\item{\textbf{\EC{} emulators are constructed for $A=4,6$ systems
      using $M$-scheme NCSM eigenstates as training data}.
  We construct emulators for \nuc{4,6}{He} and \nuc{6}{Li} in
  different model spaces from $\Nmax=4$ up to $\Nmax=10$. After the
  training phase, the emulators provide a computational speedup
  reaching seven orders of magnitude with very high output
  accuracy. In fact, we demonstrate that emulated binding energies are
  accurate to within $\lesssim 10^{-7}$ relative error in a large
  $\cD-\cE$ parameter domain with only 8--16 training points.
  }
\item{\textbf{This work demonstrates a Bayesian approach for
      handling all relevant sources of 
      uncertainty in many-body nuclear structure calculations.}
    We consider uncertainties in (i) the parametrization of the nuclear
    interaction, (ii) the model discrepancy arising from the
    truncation of the \chiEFT{} expansion, and (iii) the solution of
    the many-body problem. This study is made possible by
    employing Bayesian methods and using \EC{} emulator
    technology and the newly developed and efficient \JNCSM{}
    many-body solver. 
  }
\item{\textbf{Realistic and statistically rigorous constraints from
      few-nucleon observables lead to quantifiable propagated
      uncertainties in $A=6$ systems}.
    We employ Bayesian constraints on chiral three-nucleon forces from
    few-body observables as quantified by
    \textcite{Wesolowski:2021cni}. The parameter posterior \PDF{} from
    that study is characterized by a strong correlation and rather
    heavy tails as a result of a full treatment of EFT truncation
    uncertainties. These features are also reflected in the propagated
    parameter uncertainty for $A=6$ observables. Although our study of
    parameter uncertainty is limited to \cD{} and \cE{}---parameters
    of the leading three-nucleon force---we have hints in \nuc{4}{He}
    that propagated uncertainties from \piN{} and \NN{} LECs are small
    in comparison.
  }
\item{\textbf{A possible separation of modes in the posterior predictive
      distribution for finite nuclei is observed when using an alternative
      force calibration}.
    It is well known that correlated calibration observables lead to
    insufficient constraints on model parameters. Using, e.g., binding
    energies of \nuc{3}{H} and \nuc{4}{He} to calibrate the \cD{},
    \cE{} parameters of the \threeNF{} gives an effective constraint
    in just a single direction. We find that binding energies of $A=6$
    systems are also correlated---such that they would not offer a strong
    complimentary constraint---but there are indications for a
    separation into two modes implying that effects in heavier nuclei
    might be large.
  }
\item{\textbf{The inclusion of an EFT model discrepancy term is
      important for proper uncertainty quantification.}
    Observed order-by-order predictions corroborates a view of
    converging nuclear structure observables and allows to construct a
    statistical model for the discrepancy originating from the
    truncation of the EFT expansion. Some inferred hyperparameters of this
    statistical model---the expansion parameter $Q$ and the magnitude of
    observable coefficients $\bar{c}$---are consistent with previous
    works. In the end we employ a simplified model with fixed
    hyperparameters and recommend further statistical analysis to
    learn about the EFT convergence pattern.
  }
\item{\textbf{The treatment of correlated errors is needed for proper
      estimates of uncertainties in separation energies.}
    We find compelling evidence that order-by-order contributions to
    the binding energies in $A=4,6$ nuclei are strongly
    correlated. Assuming a single correlation coefficient to describe
    the observed convergence pattern we infer the most likely value
    $\rho=0.9$ which is then used in the uncertainty analysis. As a
    result, the final uncertainty in separation energies is better
    estimated. The uncertainties of these observables would have been
    severly overestimated without accounting for correlated errors.
  }
\item{\textbf{Method errors are the dominating source of uncertainty
      in this study}.
    The uncertainty of the many-body solution could be reduced by
    extending the model space beyond $\Nmax=10(8)$ for
    \nuc{6}{Li}(\nuc{6}{He}). We note that computations up to
    $\Nmax = 22$ have been performed for \nuc{6}{Li} resulting in
    energy convergence at the level of 10
    keV~\cite{forssen2017}. However, three-body force approximation
    schemes (with proper uncertainty quantification) will be needed to
    reach such large model spaces.%
  }
\item{\textbf{Our study provides hints for systematic underbinding in
      $A=6$ systems}.
    The observed underbinding is just at the edge of the 68\%
    credible regions for both \nuc{6}{Li} and \nuc{6}{He} implying
    very weak evidence for possible physics interpretations of this
    finding. Futhermore, the
    \Nmax{}-extrapolations that are employed in this work 
    systematically overestimate the energy convergence rate---an
    observation that we have tried to take into account in our
    statistical error model. Still, it might
    be relevant to use infrared extrapolation
    techniques~\cite{coon2012,furnstahl2012,wendt2015,forssen2017}
    to achieve a more systematic treatment and better understanding of method uncertainties.%
  }
\end{itemize}
%
%

\begin{acknowledgments}
  We thank P. Navr\'atil for useful discussions and for making it
  possible to perform validation of results from \JNCSM{} and \NCSD{}.
  This work was supported by the Swedish Research Council, Grant
  No. 2017-04234 (TD, CF) and the European Research Council (ERC)
  European Unions Horizon 2020 research and innovation programme,
  Grant agreement No. 758027 (AE). Parts of the computations were
  enabled by resources provided by the Swedish National Infrastructure
  for Computing (SNIC) at Chalmers Centre for Computational Science
  and Engineering (C3SE) and the National Supercomputer Centre (NSC)
  partially funded by the Swedish Research Council.
\end{acknowledgments}


\bibliography{master,temp,bayesian_refs}

\end{document}